\documentclass[superscriptaddress,twocolumn,pra]{revtex4-1}

\usepackage{graphicx}
\usepackage{amsmath}
\usepackage{longtable}
\usepackage{amssymb}
\usepackage{latexsym}
\usepackage{natbib}
\usepackage{color}

\usepackage{times}
\usepackage{graphics}
\usepackage{epsfig}
\usepackage{wrapfig}
\usepackage[T1]{fontenc}
\usepackage{float}
\usepackage{enumerate}
\usepackage[compact]{titlesec}
\usepackage{setspace,ifthen}
\usepackage{amsthm} 
\usepackage{amssymb}	
\usepackage{times}
\usepackage[usenames,dvipsnames]{xcolor}
\usepackage{mathrsfs,amsmath}
\usepackage{mathrsfs}
\usepackage{mathrsfs}
\usepackage{rotating}
\usepackage{epstopdf}
\usepackage{appendix}

\usepackage{csquotes}
\usepackage{bbold}
\usepackage{datetime}
\usepackage{soul}

\newcommand{\ket}[1]{\left|{#1}\right\rangle}
\usepackage{array}

\usepackage{rotate}

\newcommand{\sigI}{\hat{\sigma}_I}
\newcommand{\sigx}{\hat{\sigma}_x}
\newcommand{\sigy}{\hat{\sigma}_y}
\newcommand{\sigz}{\hat{\sigma}_z}

\newcommand{\Identity}{\hat{\mathbb{I}}}

\newcommand{\rhat}{\hat{\boldsymbol{r}}}
\newcommand{\xhat}{\hat{\boldsymbol{x}}}
\newcommand{\yhat}{\hat{\boldsymbol{y}}}
\newcommand{\zhat}{\hat{\boldsymbol{z}}}

\newcommand{\Rvec}{\vec{\boldsymbol{R}}}
\newcommand{\Vvec}{\vec{\boldsymbol{V}}}

\newcommand{\Infidelity}{\mathcal{I}}

\begin{document}
\title{Experimental quantum verification in the presence of temporally correlated noise}
\author{S. Mavadia}
\altaffiliation{\emph{These three authors contributed equally to this work.}}
\affiliation{ARC Centre for Engineered Quantum Systems, School of Physics, The University of Sydney, NSW Australia}
\affiliation{National Measurement Institute, West Lindfield NSW 2070 Australia}

\author{C. L. Edmunds}
\altaffiliation{\emph{These three authors contributed equally to this work.}}
\affiliation{ARC Centre for Engineered Quantum Systems, School of Physics, The University of Sydney, NSW Australia}
\affiliation{National Measurement Institute, West Lindfield NSW 2070 Australia}

\author{C. Hempel}
\altaffiliation{\emph{These three authors contributed equally to this work.}}
\affiliation{ARC Centre for Engineered Quantum Systems, School of Physics, The University of Sydney, NSW Australia}
\affiliation{National Measurement Institute, West Lindfield NSW 2070 Australia}
\author{H. Ball}
\affiliation{ARC Centre for Engineered Quantum Systems, School of Physics, The University of Sydney, NSW Australia}
\author{F. Roy}
\affiliation{ARC Centre for Engineered Quantum Systems, School of Physics, The University of Sydney, NSW Australia}
\author{T. M. Stace}
\affiliation{ARC Centre for Engineered Quantum Systems, School of Physics and Mathematics, The University of Queensland, St Lucia, QLD Australia}
\author{M. J. Biercuk}
\altaffiliation{Corresponding author: michael.biercuk@sydney.edu.au }
\affiliation{ARC Centre for Engineered Quantum Systems, School of Physics, The University of Sydney, NSW Australia}
\affiliation{National Measurement Institute, West Lindfield NSW 2070 Australia}
\date{\today~at~\currenttime}

\begin{abstract}
Growth in the complexity and capabilities of quantum information hardware mandates access to practical techniques for performance verification that function under realistic laboratory conditions.  Here we experimentally characterise the impact of common temporally correlated noise processes on both randomised benchmarking (RB) and gate-set tomography (GST).  We study these using an analytic toolkit based on a formalism mapping noise to errors for arbitrary sequences of unitary operations. This analysis highlights the role of sequence structure in enhancing or suppressing the sensitivity of quantum verification protocols to either slowly or rapidly varying noise, which we treat in the limiting cases of quasi-DC miscalibration and white noise power spectra. We perform experiments with a single trapped $^{171}$Yb$^{+}$ ion as a qubit and inject engineered noise ($\propto \sigz$) to probe protocol performance.  Experiments on RB validate predictions that the distribution of measured fidelities over sequences is described by a gamma distribution varying between approximately Gaussian for rapidly varying noise, and a broad, highly skewed distribution for the slowly varying case. Similarly we find a strong gate set dependence of GST in the presence of correlated errors, leading to significant deviations between estimated and calculated diamond distances in the presence of correlated $\sigz$ errors.  Numerical simulations demonstrate that expansion of the gate set to include negative rotations can suppress these discrepancies and increase reported diamond distances by orders of magnitude for the same error processes.  Similar effects do not occur for correlated $\sigx$ or $\sigy$ errors or rapidly varying noise processes, highlighting the critical interplay of selected gate set and the gauge optimisation process on the meaning of the reported diamond norm in correlated noise environments.

\end{abstract}

\maketitle

Quantum characterisation, validation, and verification (QCVV) techniques are broadly used in the quantum information community in order to evaluate the performance of experimental hardware.  A variety of techniques have emerged including randomised benchmarking (RB) \cite{Emerson:2005, Knill:2008}, purity benchmarking \cite{Wallman:2015}, process tomography \cite{Poyatos:1997, Chuang:1997, Holtzafel2015, Flammia2012}, adaptive methods~\cite{Flammia_adaptive}, and gate-set tomography (GST) \cite{Merkel:2013, BlumeKohout:2017}.  Each protocol has relative strengths and weaknesses; for instance, RB has low experimental overhead but only provides average information about gate performance, while process tomography provides more information at the cost of unfavourable scaling in measurement overhead~\cite{Blatt_QPT}. Despite their differences, these protocols share the common theme that they were originally developed and mathematically formalised assuming that error processes are statistically independent and do not exhibit strong correlations in time~\cite{Emerson:2005, Knill:2008, BlumeKohout:2017}. 

Even in highly controlled laboratory environments there are a range of noise sources that, when applied to a qubit concurrent with logical gate operations, produce effective error models that diverge significantly from the assumptions underlying most QCVV protocols.  For example, slow variations in ambient magnetic fields or drifts in amplifier gain can produce temporally correlated noise processes, often characterised through a power spectral density possessing large weight at low frequencies~\cite{Rutman, Hooge_1onf, Clarke2004}.  Moreover, these error processes may exhibit gate-dependent behavior.  So far such processes have been largely ignored in experimental QCVV, with predominantly phenomenological attempts used to explain deviations from ideal outputs~\cite{Dzurak_RB}.  Understanding that such an approach is untenable when attempting to rigorously compare QCVV results to metrics relevant to quantum error correction has recently led to an expansion of theoretical activity in this space~\cite{Flammia_RBConfidence, Flammia_unitarity, Ball:2016, Flammia_FT, Fong2017,Wallman2017}.

In this work our objectives are to experimentally characterise and explain the impact of temporally correlated noise processes on the outputs of QCVV protocols, and to identify potential modifications enabling users to improve the utility of the information returned.  We perform QCVV experiments using a single trapped $^{171}$Yb$^{+}$ ion as a long-lived, high-stability qubit. Our study implements engineered frequency noise ($\propto \sigz$) in the control system in order to study the impact of different temporal noise correlations on QCVV results. We apply noise in the two extremes, either quasi-DC offsets or noise with an effective white power spectrum to approximate slowly and rapidly varying noise, respectively. Measurements reveal that QCVV outputs diverge significantly when subject to these different types of noise, highlighting potential circumstances where the information extracted from a given protocol may no longer accurately represent the true error processes experienced by individual gates. Our experiments are compared against analytic calculations linking the underlying structure of the QCVV sequences with the manifestation of specific characteristics associated with the presence of noise correlations.

\begin{figure*}[t] 
	\centering
	\includegraphics[width=15cm]{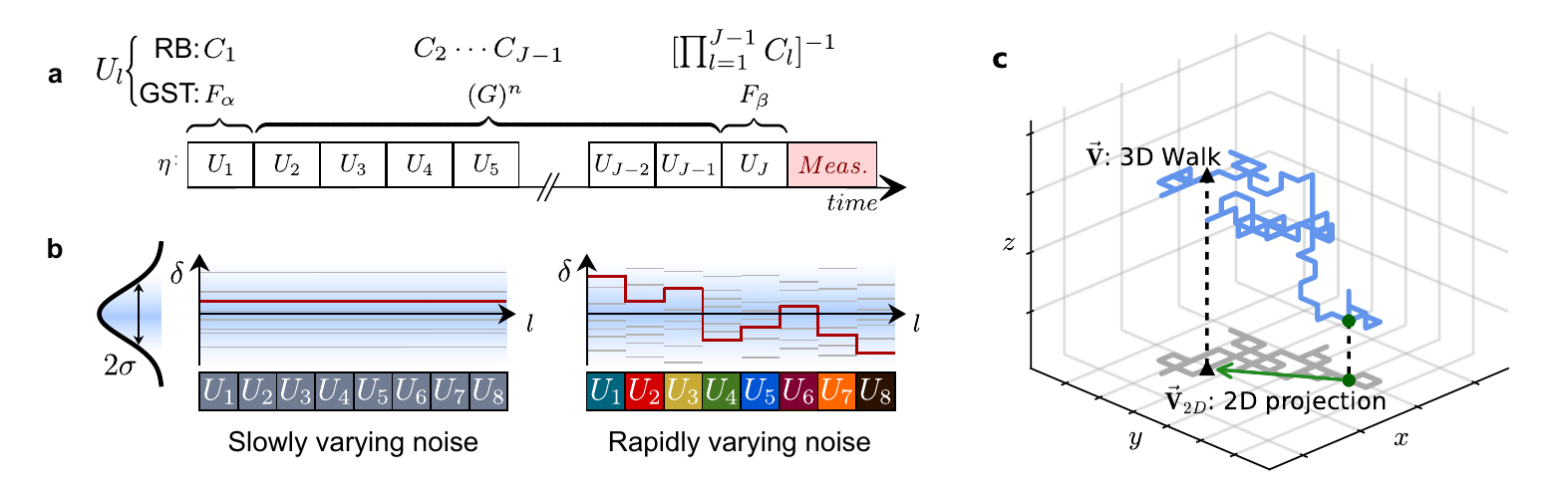} 
	\caption{QCVV sequence construction and mapping to accumulated error.  \textbf{a} Overview of unitary sequence construction for RB and GST, using Clifford gates, $C_{l}$ or fiducial operations, $F_{\alpha, \beta}$ and repeated germs $(G)^{n}$ respectively.  \textbf{b} Schematic representation of slowly and rapidly varying noise with relevant time scales defined by the sequence where $\delta$ represents the instantaneous noise values drawn from a normal distribution with $\sigma^2$ variance. Grey lines are other possible noise realisations. For RB, the noise is sampled from this distribution and varies shot-to-shot between noise realisations, while in GST a single value is selected for the entire set of experiments.  \textbf{c} Sequence-dependent ``random walk'' calculated for an arbitrary QCVV sequence (here according to the RB prescription) with $J=100$ in Pauli space. Green dot indicates origin and black triangle indicates sequence terminus.  Blue line represents the 3D walk, which can be used to calculate the trace infidelity while grey represents the 2D projection, and is measurable in a standard projective measurement. The green arrow indicates the net walk vector, $\Vvec_{2D}$, given unit step size.}
	\label{Fig:F1}
\end{figure*}

We examine two common QCVV protocols in the experimental quantum information community: RB and GST. The construction of these protocols follows a similar pattern, a series of unitary quantum operations is applied to one or more qubits sequentially in time, followed by a projective measurement (Fig.~\ref{Fig:F1}a). Experimental measurements are acquired and combined, experimental parameters are changed according to some prescription (e.g.~changing the sequence length, $J$) and further data are collected. The variation in QCVV protocols predominantly comes from the different constituent operations that are applied and the analysis techniques by which measurement results are post-processed to extract information.  

In RB, sequences are constructed by concatenating unitary operations $U_{l}$ selected at random from the 24 Clifford operations $C_{l}$.  The final operation in a sequence of length $J$ is selected to invert the net rotation $U_{J}=(\prod_{l=1}^{J-1}C_{l})^{-1}$, such that the sequence implements a net identity $\prod_{l=1}^{J}C_{l}=\Identity$.  In GST, by contrast, operations are selected deterministically according to a tabulated routine comprising specifically crafted sequences that are designed to maximise overall sensitivity to all detectable error types. These operations are constructed by concatenating so-called ``germs'', short sequences implementing predefined unitary rotations, which, in our case, are constructed from a subset of Clifford gates. The first and last unitaries $U_{1,J}\in \{F_{\alpha}, F_{\beta}\}$, termed the ``fiducial'' operations, effectively set the reference frame for state-preparation and measurement (Fig.~\ref{Fig:F1}a), see \emph{Methods} for further detail.

In our experiments we engineer noise in order to permit quantitative analysis of QCVV outputs under known conditions.  We compare the outputs obtained from both RB and GST for two distinct noise-correlation regimes. Firstly where the engineered noise is implemented as a quasi-DC miscalibration over the entire sequence, which is the extreme case for slowly varying noise and produces temporally correlated errors.  Secondly, where the engineered noise is rapidly varying (yielding an approximately white power spectrum), which leads to errors that are uncorrelated between gates (Fig.~\ref{Fig:F1}b). We now introduce a framework for interpreting the impact of sequence structure and noise correlations on measurement outcomes to facilitate an analysis of our results.

\section{Results}

\subsection{Mapping noise to measured error in RB}
The key analytic tool for our study is a formalism mapping an applied noise model to an output error for a given Clifford sequence, following a procedure derived in \cite{Ball:2016}.  Error accumulation over a given Clifford sequence maps to a ``random walk'' in a three-dimensional vector-space representing the action of sequential error unitaries in the operator space spanned by the Pauli operators, $\hat{\sigma}_{\{x,y,z\}}$ (Fig.~\ref{Fig:F1}c). For $\sigz$ noise, the $l$\textsuperscript{th} step of the walk is calculated by conjugating $\sigz$ with the entire operator subsequence $K_{l-1}\equiv\prod_{q=1}^{l-1}U_{q}$ up to the $(l-1)$\textsuperscript{th} gate, with multiplication performed from the left. This conjugation always results in a member of the Pauli group, allowing us to compactly write $\mathbf{P}_l\equiv K_{l-1}^{\dagger} \sigz K_{l-1} = \rhat_l\cdot\vec{\boldsymbol{\sigma}}$, where $\vec{\boldsymbol{\sigma}} = (\sigx,\sigy,\sigz)$ and $\rhat_{l}\in\{\pm\xhat,\pm\yhat,\pm\zhat\}$. The direction of $\mathbf{P}_l$ in Pauli space therefore maps to the Cartesian unit vector $\rhat_l$ associated with the $l$\textsuperscript{th} step of a $J$-step walk $\Rvec \equiv \sum^{J}_{l=1} \delta_{l} \rhat_{l}$. For our chosen error model, the step length, $\delta_l$, captures the integrated phase between the driving field and qubit during execution of the single gate $U_{l}$. In terms of experimental parameters, $\delta_{l} =  \Delta / \Omega$, where 
$\Delta/\Omega$ is the detuning expressed in terms of the experimental Rabi frequency, $\Omega$ (see \emph{Methods}). 

The overall form of the walk is a statistical measure of how the sequence itself interacts with the noise process to produce a net, measurable accumulation of error. Sequences that are highly susceptible to error accumulation produce walks that migrate far from the origin, while sequences exhibiting error suppression produce walks that meander back towards the origin. The net walk length is captured in the mean-squared distance from the origin $\langle \|\Rvec\| ^{2} \rangle$, averaged over noise realisations. This links to the ``trace fidelity'', defined as $\mathcal{F}_{\mathrm{trace}} = \langle | \mathrm{Tr} (\prod_{l=1}^{J} \tilde{U}_{l})| ^{2} \rangle /4$, where $\tilde{U}$ are modified unitary operations to take into account the effect of the $\sigz$ noise. We then define the infidelity $\Infidelity_{\mathrm{trace}} = 1 - \mathcal{F}_{\mathrm{trace}} \simeq \langle \|\Rvec\| ^{2} \rangle$.
 
Appropriately linking this picture of error accumulation to standard laboratory measurements requires consideration of the measurement routine itself. In typical measurements the qubit Bloch vector at the end of the sequence is projected onto the quantisation axis, $z$, with basis states $\vert 0 \rangle$ and $\vert 1 \rangle$.  A measurement of this type is therefore insensitive to net rotations around that axis of the Bloch sphere, meaning that it only probes a 2D projection of the 3D walk onto the \emph{xy}-plane.  Our preferred metric is the survival probability, $\mathcal{F}_{\textrm{survival}}$, that may be linked directly to such a 2D projection (grey line, Fig.~\ref{Fig:F1}c) as $\Infidelity_{\textrm{survival}} = 1-\mathcal{F_{\textrm{survival}}}= \langle \| \Rvec_{2D} \| ^{2} \rangle$ where $\mathcal{F_{\textrm{survival}}} = \langle | \langle 0 \vert \prod_{l=1}^{J} U_{l} \vert 0 \rangle |^2 \rangle$ , $ \langle \|\Rvec_{2D}\| ^{2} \rangle =  \langle \|\Rvec\| ^{2} \rangle -  \langle \|\Rvec_{z}\| ^{2} \rangle$, and $\langle \|\Rvec_{z}\| ^{2} \rangle$ is the mean-squared walk length along the quantisation axis (see \emph{Supplementary Material} for details). As all of our measurements are simply of the survival probability, we henceforth drop the subscripts for $\mathcal{F}$ and $\Infidelity$.  

At this stage we must link the correlation properties of the noise to the form of the walk for a specific sequence.  Considering only the underlying properties of the sequence, we may assume unit-length steps, resulting in a deterministic sequence-dependent walk with length $\Vvec~\equiv~\sum_{l=1}^{J} \rhat_{l}$.  The presence or absence of temporal noise correlations is now captured through a rescaling of the individual steps in the deterministic walk for a specific sequence. In the case of slowly varying noise, and to first-order approximation, the net error can be separated into two independent parts, {$\| \Rvec \|^2 = \delta^2 \| \Vvec \|^2$}, where $\delta$ is the value of the noise and $\| \Vvec \|$ is the net unit-step walk specific to a particular sequence~\cite{Ball:2016}. However, in the case of rapidly varying noise these two terms are no longer separable and the net error must be calculated as the convolution of the noise value at each timestep and each individual step in the random walk, $\| \Rvec \|^2 = \| \sum_{l=1}^{J} \delta_{l} \rhat_{l} \|^2$.

\subsection{Experimental platform and engineered noise}

We perform experiments using the hyperfine qubit in a single trapped $^{171}\textrm{Yb}^{+}$ ion driven by microwaves near 12.64~GHz, with basis states \mbox{$\ket{0} \equiv \ ^{2}\mathrm{S}_{1/2}\left|F=0, m_{\mathrm{F}}=0 \right\rangle$} and \mbox{$\ket{1} \equiv \ ^{2}\mathrm{S}_{1/2}\left|F=1, m_{\mathrm{F}}=0 \right\rangle$}.  Our calibration process permits accurate determination of the (first-order magnetic-field-insensitive) qubit transition frequency to within approximately $1$\,Hz.  In our laboratory, this qubit and the  associated control system have been demonstrated to possess a coherence time of $T_{2}\sim1$ s,  measurement fidelity of $\sim99.7\%$ limited by photon collection efficiency, and error rates from intrinsic system noise of $p_{RB}\approx 6\times 10^{-5}$  using ``baseline'' RB experiments (see \emph{Supplementary Figures}).  Details of the control system and experimental protocols for QCVV techniques used here are presented in the \emph{Methods}, and information about various detection procedures in use for estimating $\mathcal{F}_{\textrm{survival}}$ (including a Bayesian method) are found in the \emph{Supplementary Materials}.

We engineer $\sigz$ noise applied concurrently with Clifford operations through the application of a detuning, $\Delta$, of the qubit driving field from resonance using an externally modulated vector signal generator (see \emph{Methods}). As the detuning is applied concurrently with driven qubit rotations about $x$ and $y$ axes, rotation errors arise along multiple directions on the Bloch sphere, rather than being purely $\sigz$ in character. An additional violation of typical assumptions employed in RB is that different Clifford gates are physically decomposed into base rotations with different durations, which means that our formal error model will also be gate-dependent~\cite{Wallman2017}. 

For each of our two limiting noise cases we engineer $N$ different noise ``realisations'' in order to average over an appropriate ensemble.  In our experiments we set the distribution of noise $\Delta/\Omega \sim \mathcal{N}(0,\sigma^{2})$, where $\sigma^{2}$ is the variance of the distribution, such that the root-mean-square value is approximately equivalent in both cases once averaged over all noise realisations.  The specific implementation of noise engineering and its impact on the conduct of RB and GST is described in the \emph{Methods}, and additional details on the error model are provided in the \emph{Supplementary Materials}.

Experiments involve state preparation in the \mbox{$\ket{0}$} state, application of a unitary sequence appropriate for a QCVV protocol while subject to noise, and projective measurement of the qubit along the quantisation axis. The sequence of operations applied and the measurement procedure are determined by the protocol in use.

\subsection{RB fidelity distributions}
\begin{figure}[bp]
	\centering
\includegraphics[width=8.6cm]{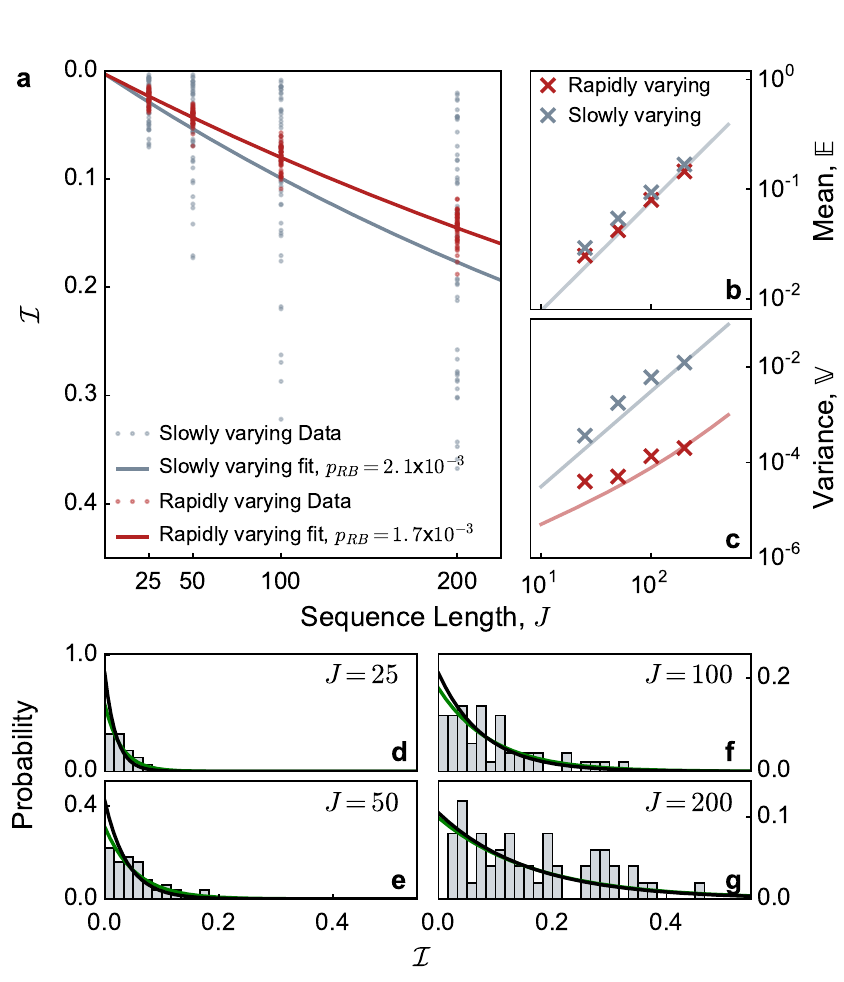} 
\caption{RB distributions over sequences in the presence of different noise correlations. \textbf{a} Standard RB protocol showing fidelity as a function of $J$ for the same set of sequences implemented under slowly varying (grey) or rapidly varying (red) noise with $\Delta_{\textrm{RMS}}=1$\,kHz.  In these experiments the Rabi frequency, $\Omega = 22.5$\,kHz. Each experiment is repeated $r=25$ to $r=30$ times under fixed conditions, and each sequence fidelity is averaged over 200 noise realisations. Lines represent exponential fits to the sequence-averaged infidelity $\overline{\mathcal{I}}=1-\overline{\mathcal{F}}(J)=0.5-(0.5-\kappa)e^{-p_{RB}J}$, weighted by the variance over sequences for each $J$, and are used to extract $p_{RB}$.  Here $\kappa=3\times10^{-3}$ represents state preparation and measurement error. \textbf{b-c}  Scaling of $\mathbb{E}(\mathcal{I})$ and $\mathbb{V}(\mathcal{I})$ against sequence length $J$, comparing experimental values (markers) against first-principles theory (lines) as per~\cite{Ball:2016} modified to state fidelity (2D walk) and noise applied concurrently with gate implementation. See \emph{Supplementary Materials} for details. \mbox{\textbf{d-g}} Histograms for data in panel \textbf{a} in the presence of slowly varying noise.  Green line: fitted gamma distribution with shape parameter fixed, $\alpha =1$.  Black line: gamma distribution using input parameters calculated from first principles (see text). $\chi^{2}$ values for calculated (fitted) gamma distributions are $\{0.354 (0.091), 0.212 (0.078), 0.241 (0.204), 0.348 (0.348)\}$.} 
\label{Fig:F2}
\end{figure}


 In the limit of rapidly varying noise, all sequences of randomly ordered Clifford gates with length $J$ are equivalent under noise averaging, and all sequence survival probabilities tend towards the mean. Recent theoretical studies have demonstrated that measurements on RB sequences in the presence of temporal noise correlations, can produce a divergence between average and worst-case reported trace fidelities~\cite{Ball:2016, Kueng:2016}.  Thus we find that measurement outcomes for different RB sequences are characterised by distributions with distinctly different shapes depending on the temporal correlations in the noise.  The standard practice of combining all measurements to extract an RB error rate, $p_{RB}$, from the decay of the mean over all $J$-gate sequences as a function of $J$, results in a global ensemble average and does not take advantage of this information (formally, as the noise we implement exhibits temporal correlations, the value of $p_{RB}$ one extracts may not be meaningful as a measure of average Clifford gate error). Our analysis takes advantage of the additional information which is always present in a RB experiment in order to evaluate the impact of noise correlations and deduce useful information about the underlying error process.  

In our experimental study we measure the noise-averaged survival probabilities for a set of sequences $\{\eta_{i}\}_{J}$, indexed by $i$ and of length $J$, for different lengths \mbox{$25 \leq J \leq 200$} (Fig.~\ref{Fig:F2}a), where we implement the same set of $J$-gate sequences under application of either slowly or rapidly varying detuning noise. For an arbitrary individual sequence, $\eta_{i}$ and a single noise realisation, $n$, we perform $r$ nominally identical repetitions of the experiment. We combine the information from the outcomes of these individual repetitions to produce a maximum-likelihood estimate of survival probability, $\mathcal{F}_{i,n}$ (see \emph{Supplementary Materials}).  The use of multiple repetitions under identical conditions reduces quantum projection noise in the qubit measurement and assists in isolating specific quantitative contributions to the distribution of survival probabilities, though this is not possible without noise engineering. In general, we average measured outcomes over a fixed number of noise realisations to yield $\mathcal{F}_{i, \langle\cdot\rangle}$ for a fixed sequence $\eta_{i}$.  From here on, we will refer to this noise-averaged survival probability as $\mathcal{F}$.

In the case of rapidly varying noise we observe the distribution of sequence outcomes is symmetrically spread around the sequence-averaged mean survival probability, $\overline{\mathcal{F}}(J)$, and the entire distribution shifts away from zero error with increasing $J$ (red data, Fig \ref{Fig:F2}a). The presence of slowly varying noise, by contrast, produces a broad distribution of measured $\mathcal{F}$ over each set $\{\eta_{i}\}_{J}$, demonstrating a positively skewed set of outcomes and the persistence of a long tail at higher error rates (lower survival probabilities).  In this case, as $J$ increases the distribution broadens but remains skewed.  Under both noise correlation cases, the measured $\overline{\mathcal{F}}(J)$ remain approximately the same. The differences in the distribution of measured survival probabilities over sequences under these two noise models reproduces the central predictions of Ref.~\cite{Ball:2016}.

We compare the characteristics of the distributions themselves against analytic predictions for both slowly and rapidly varying noise, beginning with the measured expectation, $\mathbb{E}(\mathcal{I})$, and variance, $\mathbb{V}(\mathcal{I})$ (Fig.~\ref{Fig:F2}b-c), finding good agreement by taking only the applied noise strength as an input into a theoretical model (see \emph{Supplementary Materials}). More specifically, theoretical predictions suggest that the distribution of outcomes under both noise models -- as well as intermediate models described by coloured power spectra -- should be well described by a gamma distribution~\cite{Ball:2016}. The general gamma distribution probability density function is given by 

\begin{equation}
	\Gamma(\alpha,\beta):\;\;f(x) = \frac{x^{\alpha - 1}}{\Gamma(\alpha) \beta^{\alpha}} \exp{\left[ -\frac{x}{\beta}\right] },
\end{equation}

\noindent where $\alpha$ and $\beta$ are the shape and scale parameters and $\Gamma(x)$ is the gamma function. The form of the gamma distribution will vary significantly between the limiting noise cases treated here, tending towards a symmetric Gaussian for rapidly varying noise and a broader positively skewed distribution in the presence of slowly varying noise, as determined by the values of $\alpha$ and $\beta$.

Figures~\ref{Fig:F2}d-g show histograms of RB sequence survival probabilities in the presence of the extreme case of slowly varying noise, quasi-DC miscalibration.  We overlay gamma distributions calculated from first principles using no free parameters (black lines) as $\Gamma(1, 2J\sigma^{2}/3 (1/2 + \pi^{2}/96))$, and fixing $\alpha=1$ while allowing $\beta$ to vary as a fit parameter (green lines). The theoretical prediction captures both the measured skew towards high survival probabilities and the approximate ``length'' of the tail at low survival probabilities.  We believe that residual disagreement between data and first-principles calculations arises due to both limited sequence sampling and contributions from higher-order analytic error terms when the approximation $J\sigma^{2} \ll 1$ is no longer valid.  Importantly, data and theory show the mode of the distribution is close to unity survival probability ($\mathbb{I}=0$) and therefore corresponds to a lower error than the mean.  For details on modifications to the theory presented in~\cite{Ball:2016} accounting for the specific noise and gate-dependent error model employed in our experiments, contributions from higher-order terms, and expanded data sets including larger sequence numbers, see \emph{Supplementary Material}.

\subsection{Modification of RB for identification of model violation}
\begin{figure}[t] 
   \centering
   \includegraphics[width=\columnwidth]{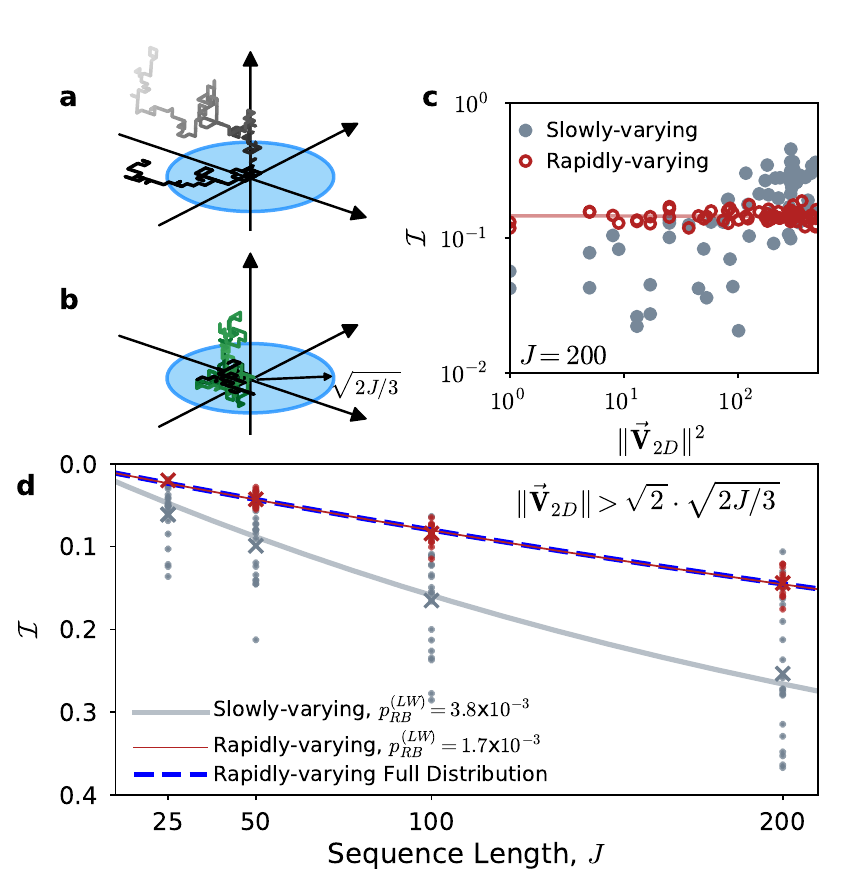} 
   \caption{RB using long-walk sequences.  \textbf{a-b}) Schematic representations of long (a) and short (b) length walks under slowly varying noise in 3D (coloured lines) and 2D (black lines), defined relative to a limit deduced from diffusive behaviour, as indicated by the blue circle. \textbf{c}) Noise-averaged fidelity distributions of the same sequences as a function of walk length in the 2D plane. Measured infidelity vs.~2D walk length, $\|\Vvec_{2D}\|^{2}$, when subject to slowly varying (grey) and rapidly varying (red) noise with linear fit overlaid. The slope of this fit is $(0.8 \pm 1) \times 10^{-5}$, consistent with zero. \textbf{d}) RB using long-walk sequences. Solid red line corresponds to RB performed using 20 long-walk sequences and rapidly varying noise.  Extracted $p^{(LW)}_{RB}$ matches that extracted under the same conditions using an unbiased sampling of all sequences (dashed line). Grey line corresponds to RB using the same long-walk sequences and slowly varying noise.  For the exponential fits, state-preparation and measurement error, $\kappa$, is fixed to $3\times 10^{-3}$.}
   \label{Fig:F3}
\end{figure}

The fact that the distribution of sequence survival probabilities under slowly varying noise does not converge to the mean indicates sequence-dependence in the resulting error accumulation.  The emergence of this phenomenology is elucidated through an examination of the walks for different sequences.  Under this type of noise certain sequences possess walks with large $\|\Vvec_{2D}\|^{2}$, hence amplifying the accumulation of error, while others tend back towards the origin and show reduced accumulated error (Fig.~\ref{Fig:F3}a-b). We arbitrarily classify sequences as ``long-walk'' if they possess a 2D projection beyond the diffusive mean-squared limit for an unbiased random walk, $\|\Vvec_{2D}\|^{2}> \frac{2}{3}J$.

We link between the sequence walk in Pauli space and the noise-averaged survival probability by displaying the experimentally measured $\Infidelity$ for sequences of fixed length $J=200$ against the calculated 2D walk length, $\|\Vvec_{2D}\|^{2}$ (Fig.~\ref{Fig:F3}c).  Data are presented for both rapidly varying (red open markers) and slowly varying (grey solid markers) noise, where the same set of sequences is used between the noise models.  Measurements for rapidly varying noise are fit with a line possessing slope approximately consistent with zero, while for the same sequences under slowly varying noise, the measurements show a positive dependence on $\|\Vvec_{2D}\|^{2}$ as expected. We believe the significant scatter in the plot is partially due to a concurrently acting noise source and higher-order contributions to error, neither of which are incorporated in the first principles calculation of the walk, $\|\Vvec_{2D}\|^{2}$ (see \emph{Supplementary Materials} and Appendix \textbf{C} of~\cite{Ball:2016}).  Nonetheless, the effect of sequence structure on measured survival probability is clearly visible for the case of slowly varying noise.

In aggregate, this phenomenology gives rise to the skewed gamma distribution under slowly varying noise described above, and the convergence of all noise averaged survival probabilities for individual sequences to the ensemble average when the noise is rapidly varying.  However, pre-selection of RB sequences possessing large calculated, unit-step walks also provides a mechanism to both identify the presence of temporally correlated errors and extract an RB outcome that more closely approximates worst-case errors.  In Fig.~\ref{Fig:F3}d we plot $\Infidelity$ vs.~$J$ for a subset of sequences preselected to possess long walks as in Fig~\ref{Fig:F3}a-b, whose survival probabilities we denote $\mathcal{F}_{LW}(J)$.  We choose that the preselection of long walks is based on the condition $\|\Vvec_{2D}\|^{2}> 2\times \frac{2}{3}J$.
 
When these long-walk sequences are subjected to rapidly varying noise, the distribution of survival probabilities over sequences remains approximately Gaussian about the mean, and the expectation value over this subset closely approximates the expectation value over an unbiased random sampling of the $24^{J}$ possible $J$-gate sequences, $\overline{\mathcal{F}}_{LW}^{\mathrm{rapid}}\approx\overline{\mathcal{F}}^{\mathrm{rapid}}$, (Fig. \ref{Fig:F3}d, red solid line and blue dashed line).  However, in the presence of slowly varying noise we observe a larger error rate than that achieved with unbiased sampling \mbox{$\overline{\mathcal{F}}_{LW}^{\mathrm{slow}}>\overline{\mathcal{F}}^{\mathrm{slow}}$}.  The difference between the sequence-averaged expectation values in these noise cases arises solely because of the intrinsic properties of the sequences in use.

Extracting a RB gate-error-rate, $p^{(LW)}_{RB}$ from $\overline{\mathcal{F}}_{LW}(J)$ in the presence of slowly varying noise, we typically find an increase $p^{(LW)}_{RB}\sim 2 - 5\times p_{RB}$  relative to standard sequence sampling, depending on the number of long-walk sequences employed, and the threshold value of $\|\Vvec_{2D}\|^{2}$ used to define a ``long walk'' (Fig.~\ref{Fig:F3}c). This approach effectively constitutes construction of an RB protocol that increases the reported error rate by enhancing sensitivity to a particular noise type, which in our case is $\propto \sigz$.  Alternative sequences may also be calculated that are more sensitive to $\sigx$ or $\sigy$ noise than randomly selected RB sequences.  These error enhancing sequences give a clear, qualitative signature of the violation of the assumption that the error process is uncorrelated in time, although we do not claim that such a signature is in general uniquely associated with the presence of temporal noise correlations. Furthermore, because calculation of $\|\Vvec_{2D}\|^{2}$ and sequence pre-selection is performed numerically in advance, this approach alleviates the requirement to average extensively in experiment over sequences in order to reveal the skewed fidelity distribution.

\subsection{GST in the presence of correlated noise}
\begin{figure}[ht!] 
   \centering
   \includegraphics[width=8.8cm]{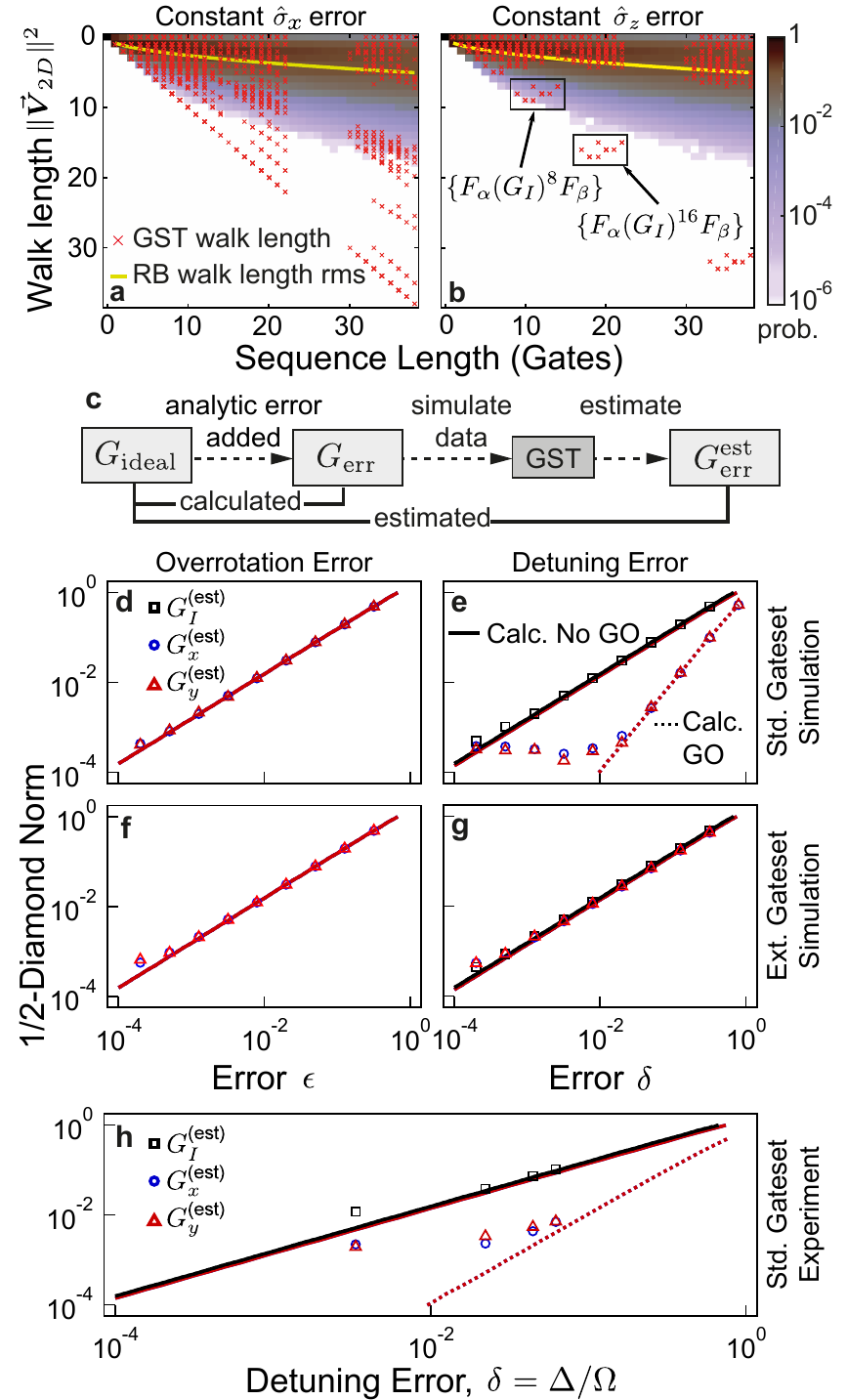} 
   \caption{Demonstration of GST sensitivity to correlated error models.  \textbf{a, b)} Sensitivity of GST sequences to $\sigx$, $\sigz$ errors using the length of the sequence-dependent walk vector $\Vvec_{2D}$. GST sequence walks are shown as red crosses on a background colour scale illustrating the distribution over $10^6$ RB walks and their average (yellow line). Here "gates" are defined as constituent Clifford operations of length $\tau_{\pi/2}$.
 \textbf{c)} Flow diagram for the numerical analysis of the diamond norm estimation under correlated errors concurrent with gates $G$. 
 \textbf{d, e)} Results of the analysis for the standard gate set ${G_I, G_x, G_y}$ with the calculated diamond distance shown as solid lines (dashed lines) without (with) gauge optimization on all graphs, and GST estimation depicted as symbols. Both overrotation errors on the $G_x, G_y$ gates (d) and concurrent detuning errors (e) are studied. For overrotation errors the ideal rotation angle, $\theta\to(1+\epsilon)\theta$.  \textbf{f, g)} Analysis is repeated by extending the gate set to include ${-G_x, -G_y}$. In panels (d) and (f) which employ only overrotation errors, the calculated diamond distance for $G_{I}$ vanishes and we do not show the noise floor for visual clarity.  \textbf{h)} Experimental investigation of concurrent detuning $\sigz$ errors via a deliberately engineered detuning $\Delta$.
Markers indicate GST estimates from experimental data and solid lines represent analytical calculations performed within the pyGSTi toolkit.}
   \label{Fig:F4}
\end{figure}

We now apply the sequence-dependent Pauli walk framework to GST in order to understand the interplay of sequence structure and temporal noise correlations in the GST estimation procedures.  We begin by collating all standard GST sequences up to 256 gates in length using gates $G_{I}\equiv\Identity$, the identity, $G_{x}$, a $\pi/2$ $\sigx$ rotation and $G_{y}$ defined similarly.  We define sequences to include fiducial operations and germs (see \emph{Methods} and Ref~\cite{Blume-Kohout2017}), and calculate the corresponding walk lengths.  Here we assume unit step size under application of either a quasi-DC $\sigz$ or $\sigx$ unitary error process (Fig.~\ref{Fig:F4}a, b) such that $\|\Rvec\|^2 = \delta^2 \|\Vvec\|^2$, and plot $\|\Vvec_{2D}\|^{2}$ as a proxy for projected sequence error vs. $J$. We overlay the results on the calculated probability distribution of unit-step walks for RB sequences, presented as a colour scale for comparison. Points appear clumped due to the GST prescription using different fiducials (leading to different sequence lengths) surrounding a reported germ, as highlighted in Fig.\,\ref{Fig:F4}b.
 
Examining these data indicates that GST sequences broadly sample the range of expected fidelities in the presence of strongly correlated $\sigx$ errors, more effectively so than RB. However, their structure appears to systematically suppress measured errors in the presence of correlated $\sigz$ errors.  This mimics the positive skew of RB sequence survival probabilities in the presence of slowly varying noise, as observed in the colour scale. In the presence of correlated $\sigz$ errors, only GST sequences consisting of repeated $G_{I}$ germs, formally equivalent to Ramsey experiments~\cite{Ramsey1950}, show sensitivity to this kind of error.  We now explore the impact of these observations in further detail by both numerical investigations and experiments involving engineered unitary $\sigz$ errors. 

Given measurement outcomes (experimental or simulated) for the prescribed sequences, the open-source analysis package pyGSTi \cite{pyGSTi} is used to extract a large set of results characterising the performance of the gate set. One important metric calculated by the protocol for each gate is the diamond distance, $\|G_{\text{ideal}}-G\|_{\diamond}$, which is meant to provide a worst-case bound on the distance to the ideal gate operation. GST has found wide adoption in part because of its ability to calculate this metric, which is postulated to be important for formal analyses of fault-tolerance in the context of quantum error correction.

In our first test, we numerically probe the sensitivity of the GST analysis procedure to correlated error using the aforementioned pyGSTi toolkit \cite{pyGSTi}. We introduce constant $\sigx$, $\sigy$, or $\sigz$ errors via concurrent unitary rotations added to the formerly ideal operations.  
Therefore the exact mathematical representation of each gate ($G_{I,x,y}$) is known from analytical transformations and we have two paths to evaluate gate performance (Fig.~\ref{Fig:F4}c). First, we directly calculate the diamond distance ($\|G_{\text{ideal}}-G_{\text{err}}\|_{\diamond}$) using the matrix representation of $G_{\text{err}}$, maintaining the initial frame of reference. Second, we estimate it by employing pyGSTi to simulate data using $G_{\text{err}}$ and determine the diamond distance ($\|G_{\text{ideal}}-G^{(\text{est})}_{\text{err}}\|_{\diamond}$) of the estimate $G^{(\text{est})}_{\text{err}}$ obtained by the toolkit's fitting routines. 

The GST estimation procedure incorporates a gauge optimisation by construction, as it makes no assumptions in regard to the qubit and its measurement basis. It performs two rounds of gauge optimisation, allowing identification of a frame in which to minimise the distance to the entire set of target gates.  The relevance of this gauge freedom on RB-derived estimates of gate performance was highlighted recently in~\cite{GST_Gauge}. To illustrate how gauge freedom affects the results, we separately calculate the diamond distance with and without gauge optimising our analytic gate set $G_\text{err}$ using routines included in the pyGSTi toolkit.

We plot the calculated and estimated diamond norms for $G_{I,x,y}$, subject to processes similar to either a constant over-rotation (\emph{i.e.} proportional to $\sigx$ or $\sigy$ depending on the germ with no error on $G_{I}$ operations), or a constant detuning error (\emph{i.e.} proportional to $\sigz$), as shown Fig.~\ref{Fig:F4}d-e.  Here we see that while the estimated diamond distance for operators $G_{I,x,y}$ closely matches the calculated value in the presence of numerical overrotation errors, GST appears to significantly underestimate $G_{x}$ and $G_{y}$ errors arising from constant unitary $\sigz$ errors, and only the diamond norm estimate for $G_{I}$ appears similar to the directly calculated value. Other estimated quantities such as process infidelity and the associated Choi matrices are affected in a similar way (see \emph{Supplementary Material}). However, performing gauge optimisation on the analytically calculated matrices $G_\text{err}$ as well (within the pyGSTi package) reduces the reported diamond distance for $\sigz$ errors, and produces agreement with the much lower $G_{x,y}$ diamond distance reported by the GST estimation procedure (Fig.~\ref{Fig:F4}e). Among the error models we have tested, for this gate set such behavior is only manifested in the presence of temporally correlated $\sigz$ errors and does not appear using various other error processes built into the GST analysis package (see \emph{Supplementary Material} for details) .

To further investigate the influence of the gauge degree of freedom, we repeat our numerical analysis under the application of identical unitary errors, but extend the gate set by adding negative rotations $-G_x$, $-G_y$ corresponding to $-\pi/2$ $\sigx$ and $\sigy$  rotations and incorporating a number of associated compound germs (Fig.~\ref{Fig:F4}f-g).  The resulting gauge-optimised calculated and estimated diamond-distance values now increase, moving closer to the analytic calculation obtained \emph{without} gauge optimisation.  This simple change in the gate set directly reveals the role of gauge optimisation in the discrepancies we noted above, as the additional information provided to GST via the extended gate set effectively constrains the gauge optimisation procedure.

We follow up on these numerical investigations by performing experiments using GST gates subjected to engineered unitary $\sigz$-errors of varying strength.  As before, we generate an operation with known error magnitude and form, allowing us to directly produce a matrix representation for the gate and hence calculate the diamond distance for the (deliberately) imperfect gates we apply to our trapped-ion system.  Again GST produces an estimate of the diamond distance that matches the calculation for $G_{I}$, but yields estimates of the diamond distance from experimental data approximately an order of magnitude below the (unoptimised) calculated value for $G_{x,y}$ (Fig.~\ref{Fig:F4}h).  Allowing gauge optimisation on the calculated diamond distance changes its scaling with error magnitude as in simulations above. We do not find strong agreement between data for $G_{x,y}$ and this gauge-optimised scaling, but cannot exclude the possibility that other finite sampling effects may cause saturation of small reported diamond distances.  

In addition to the cases presented above we have also performed experimental GST with a wide variety of engineered, time-varying errors.  These include detuning and amplitude noise exhibiting 50 Hz fluctuations and slow drifts (\emph{i.e.} varying in time during individual sequences), constant overrotations, and added state-preparation and measurement (SPAM) errors.  All data sets, corresponding pyGSTi analysis files and resultant reports are included as part of the \emph{Supplementary Material}.


\section{Discussion}
In our studies we have employed a simple analytic framework - a formalism mapping noise to error accumulation in sequences of Clifford operations - to explore the sensitivity of RB and GST to slowly varying noise processes. Theoretical predictions derived from this framework match RB experiments employing engineered noise with known characteristics: either slowly varying or rapidly varying on the sequence timescale. This highlights the utility of the random-walk analysis in determining sequence-dependent sensitivities of QCVV protocols in the presence of temporally correlated noise.  

We have compared RB survival probabilities over sequences to a gamma distribution $\Gamma(\alpha=1, \beta)$, where $\beta$ is determined by the type of error model employed in the experiment, and shown good agreement using no free parameters. In addition we have demonstrated that in the presence of slowly varying noise, the mode of the distribution of survival probabilities over sequences is shifted towards lower error rates than the mean and that a long tail of high-error outcomes appears as predicted in \cite{Ball:2016}.  

Overall, the experiments reported here give a clear experimental signature of the violation of the assumption that errors between gates are independent. While we do not claim that the features we observe are in general uniquely derived from this interpretation, we hope these results may help experimentalists seeking to interpret complex RB data sets.   We believe that more detailed reporting of RB outcomes including the publication of distributions over $\mathcal{F}$, as well as the sequences employed, will facilitate more meaningful comparisons between RB data sets derived from different physical systems, as the relevance of $p_{RB}$ is diminished when error processes exhibit temporal correlations.

Through a combination of analytic calculations, numerics, and experiments with engineered errors we have found a similar bias towards high estimates of gate fidelity in GST (using a standard $G_{I,x,y}$ gate set) subjected to strongly correlated, unitary $\sigz$ errors. The asymmetry we observe between the manifestation of correlated $\sigx/\sigy$ and $\sigz$ error-sensitivity in GST outputs has not been reported previously, to the best of our knowledge.  We have shown explicitly how the low diamond-distance estimates under this kind of noise are related to the gauge optimisation performed as part of the protocol; limiting the gauge freedom by extending the gate set under application of an identical error process dramatically changed the estimated diamond distance of the very same gates in numerical simulations. 

These observations are commensurate with a simple physical interpretation of the effect of an optimised gauge transformation in the circumstances we examine.  In the presence of correlated $\sigz$ errors, when the gate set is limited to $G_{I,x,y}$ gates, the reconstructed operator includes an extra error component along the $z$-axis. The effect of gauge optimisation is to rotate the operators for $G_x$ and $G_y$ back to the equatorial plane, effectively cancelling this error. Under this circumstance the magnitude of rotation of these gates is smaller than expected in a fixed lab frame, and the second-order nature of the residual errors result in a steeper gradient of the dotted line in Fig.~\ref{Fig:F4}e. In contrast the $G_{I}$ rotation should have no net rotation and therefore this error will not be cancelled by a simple gauge transformation.

Gauge optimisation is designed to produce the best estimate for errors over the entire gate set, and in a sense acts to ``distribute'' nominal errors over all constituent rotations in the gate set.  The validity of such a gauge transformation in the presence of independent protocols for establishing a measurement basis remains an open question and has been highlighted recently by Rudnicki et al. \cite{Rudnicki:2017}.  The variation of calculated and estimated diamond distances under correlated $\sigz$ errors when subjected to seemingly small modifications of the gate set has again not been reported previously, and indicates an important dependence of GST output on the specific gate set employed, the characteristics of the underlying error source, and the gauge optimisation procedure.  


Clearly the observed performance of GST in the presence of correlated $\sigx$ noise, such as resulting from experimental over-rotations, can make GST a valuable tool in debugging an experimental system \cite{Dehollain:2016}, although precise calibrations can also be carried out efficiently using a subset of the full GST protocol \cite{Rudinger:2017}. The effect of gauge optimisation in the presence of $\sigz$ errors, however, is concerning as a key implied benefit of GST is its ability to directly estimate the diamond distance and hence provide a rigorous upper bound on gate errors using a fully self-contained analysis package. Recent experimental work \cite{BlumeKohout:2017} on the topic claimed such upper bounds on gate errors using GST and compared these to the fault-tolerance threshold with high reported confidence and tight uncertainties.  The results above call into question the relevance of the reported metrics without either additional independent verification and noise characterisation, or a more detailed discussion on the relationship between the gate set implemented in GST and the gates which could potentially be employed in a calculation using that system. Furthermore, when acquiring and evaluating data, care has to be taken to to suppress any form of model violations reported by the GST toolkit in its likelihood analysis, as otherwise the extracted performance metrics may become unreliable. These deviations are currently not reflected in the uncertainties (\emph{i.e.} error bars) calculated for those metrics by the toolkit and discussions with its authors suggest that a connection between the two is a non-trivial process.

In light of the investigations reported here, we believe that there is a need for greater awareness of the subtleties of the use of both RB and GST in the presence of temporally correlated noise environments.  In order to enhance the meaning and utility of reported results we advocate that QCVV benchmarks such as $p_{RB}$ and GST diamond distances should be reported together with a quantitative measure of violation from a purely Markovian, temporally uncorrelated model. In the case of RB, this could be the difference between the extracted $p_{RB}$ of long and short walk sequences; in GST the deviation is already being reported as part of the routine, yet the question about the impact of gauge optimisation that we identified remains.  Similarly, if using GST as a standalone gate evaluation procedure one cannot know a priori the form of the underlying noise - and hence any associated GST insensitivities.  Increasing the rigour of resultant upper bounds on diamond distances could require performing GST using multiple different gate sets in order to identify potential ``blind spots.''  Given the experimental overhead, however, this brute force approach is not necessarily attractive and further modifications to GST could resolve the issue with considerably greater efficiency. Overall, we hope that these observations will assist in both the interpretation of QCVV experiments when model violation may occur, and the development of new techniques with improved rigour and efficiency for larger scale systems.


\section{Methods}

\subsection*{Experimental gate implementation}
Quantum gates are implemented on a single $^{171}$Yb$^+$ ion by driving its qubit transition at $12.6$~GHz with microwave pulses produced by a vector signal generator (VSG, model Keysight E8267D). The phase of the driving field is adjustable via I-Q modulation allowing us to implement rotations around any axis lying in the $xy$-plane of the Bloch sphere. Rotations around the $z$-axis are carried out as frame-updates, \emph{i.e.} pre-calculated, instantaneous changes of the generator I-Q values. Identity operations are realised as \enquote{idle} periods, whereby no signals are applied for a time equivalent to that of a $\pi$ or $\pi/2$ rotation. We additionally implement pulse modulation (\enquote{RF blanking}) to suppress transients in microwave power at pulse edges. In this way, we implement the full set of Clifford gates as listed in supplementary materials.

All RB and GST sequences are uploaded to the VSG prior to the experiments and selected when required. When the number of implemented sequences is large, as is the case with GST, the latter step is the bottleneck in our experiments as sequence selection, depending on the constituent number of gates $J$, can take up to tens of seconds using our signal generator due to the use of the in-built, high-suppression, RF blanking switch which adds significant overhead.

\subsection*{Experimental noise implementation}
In RB experiments correlated noise is implemented by shifting the VSG drive frequency by a fixed amount based on a list of $N=200$ samples from a Gaussian noise distribution (see \emph{Supplemental Information}). The same list of noise realisations is repeated for each RB sequence in a set of given length $J$, yielding sets of noise-averaged fidelities. In GST experiments we implement constant noise of the same strength over all the sequences. Only four noise detunings are implemented due to the large overhead imposed by sequence selection prior to execution.

Rapidly varying noise in RB is implemented via the VSG's external frequency modulation, whereby the frequency offset is encoded as a series of calibrated offset voltages on an arbitrary waveform generator (Keysight 33622A) and supplied time-synchronous to each gate within a sequence. Again, $N=200$ different realisations, each consisting of $J$ samples are applied to each RB sequence to extract a noise-averaged fidelity.  Further details can be found in the \emph{Supplementary Material}.

\subsection*{Concurrent noise model and gate dependent errors}

Deliberately induced $\sigz$ errors are implemented via a fixed detuning $\Delta$ from the qubit's transition frequency, which is tracked by regularly spaced Ramsey experiments to better than 1~Hz accuracy relative to a Rabi frequency of \mbox{$\Omega =$ 22.5\,kHz}. We apply noisy gates in which a concurrent $\sigz$ rotation modifies the unitary evolution of our physically implemented gates (only \mbox{$\sigI\equiv\Identity$, $\sigx$ and $\sigy$}) given by matrix-exponentiation of the corresponding Pauli-matrices $\hat{\sigma}_{\{I,x,y\}}$ as

\begin{equation}
\label{eq:unitary_w_err}
\tilde{U}(\theta,\Delta,\Omega) = \exp \left\{-i\left[ \bigg(\frac{\theta}{2} \hat{\sigma}_{\{I,x,y\}}\bigg) + \bigg(\frac{\vert{\theta}\vert}{2} \frac{\Delta}{\Omega} \sigz \bigg)\right]  \right\}. 
\end{equation}
\noindent The first term in the exponential corresponds to the unperturbed unitary where the rotation angle $\theta$ is chosen to be either \mbox{$\theta=\pm\pi$} or \mbox{$\theta = \pm\pi/2$}.  Here the effective error magnitude scales in relation to the Rabi frequency $\Omega$, and the absolute value of $\theta$ ensures that the sign of the detuning term is preserved under positive and negative gate rotations.

This implementation leads to gate dependent errors. Hence $\pi$ rotations accumulate twice the phase i. n the presence of a nonzero $\Delta$ as $\pi/2$ rotations.

\subsection*{Gate Set Tomography}
Initial $F_\alpha$ and final $F_\beta$ fiducial operations are taken from the set $\{\emptyset, G_x, G_y, G_x G_x, G_x G_x G_x, G_y G_y G_y \}$, where $\emptyset$ stands for no gate operation, and $G_x$ and $G_y$ stand for $\pi/2$ rotations around the $x$ and $y$-axes of the Bloch sphere. They are chosen to form an informationally complete set of input states and measurement bases akin to quantum process tomography. The germs used in our experiments are \\
$G_x$, $G_y$, $G_I$, $G_x G_y$,\\ 
$G_x G_y G_I$, $G_x G_I G_y$, $G_x G_I G_I$, $G_y G_I G_I$,\\
 $G_x G_x G_I G_y$, $G_x G_y G_y G_I$,\\
 $G_x G_x G_y G_x G_y G_y$,\\
identical to those used in reference~\cite{BlumeKohout:2017} and recommended as \enquote{standard} GST in the pyGSTi tutorials.
In our numerical analysis, we extend the standard gate set from $\{G_I, G_x, G_y\} \rightarrow \{G_I, G_x, G_y, -G_x, -G_y\}$ while also significantly expanding the germ set from 11 to 39 elements (see \emph{Supplemental Material} for details).
Each of these germs is concatenated with itself up to a maximum length that successively increases as \mbox{$L=\{1, 2, 4, 8, 16, 32, 64, 128, 256\}$} and measured in all 36 combinations of the fiducials $F_\alpha$ and $F_\beta$. 
In the experimental implementation, we first record a baseline measurement without added error and then step through the cases of added detunings \mbox{$\Delta=\{75, 500, 1000, 1400\}$~Hz} for all 2737 sequences of the standard set. Due to overhead associated with switching between sequences, we recorded 220 repetitions for each sequence in consecutive order. The toolkit's authors advise to instead interleave sequences and repetitions to spread slow drifts across the data set in order to reduce model violations in the fitting routines \cite{ENielsen:2017}. 

\emph{Acknowledgements}:  
The authors acknowledge discussions with R.-Blume Kohout, P. Maunz, K. C. Young, and E. Nielsen on GST, and R. Harper, C. Ferrie, and C. Granade for discussions on data analysis. We are grateful to J. Emerson to pointing out the potential utility of adding negative rotations to the GST gate set.  Work partially supported by the ARC Centre of Excellence for Engineered Quantum Systems CE110001013, the Intelligence Advanced Research Projects Activity (IARPA) through the US Army Research Office, and a private grant from H. \& A. Harley.

\emph{Author Contributions}:
S. M. and C. E. led experimental implementation, data collection, and data analysis for RB. C. H. led experimental design, simulations, and analysis of GST.  S. M., C. E. and C. H. jointly produced the figures. H. B., F. R., and T. M. S. performed theoretical analyses and calculations.  M.J.B. conceived the general direction of this study, oversaw experimental design, and led writing of the manuscript and production of figures.  

\emph{Competing Interests}:
The authors have no competing interests.


\begin{thebibliography}{10}
\expandafter\ifx\csname url\endcsname\relax
  \def\url#1{{\tt #1}}\fi
\expandafter\ifx\csname urlprefix\endcsname\relax\def\urlprefix{URL }\fi
\providecommand{\eprint}[2][]{\url{#2}}

\bibitem{Emerson:2005}
Emerson J, Alicki R and {\.{Z}}yczkowski K 2005 {\em Journal of Optics B:
  Quantum and Semiclassical Optics\/} {\bf 7} S347--S352
  \urlprefix\url{http://iopscience.iop.org/article/10.1088/1464-4266/7/10/021}

\bibitem{Knill:2008}
Knill E, Leibfried D, Reichle R, Britton J, Blakestad R~B, Jost J~D, Langer C,
  Ozeri R, Seidelin S and Wineland D~J 2008 {\em Physical Review A\/} {\bf 77}
  012307 \urlprefix\url{http://link.aps.org/doi/10.1103/PhysRevA.77.012307}

\bibitem{Wallman:2015}
Wallman J, Granade C, Harper R and Flammia S~T 2015 {\em New Journal of
  Physics\/} {\bf 17} 113020
  \urlprefix\url{http://iopscience.iop.org/article/10.1088/1367-2630/17/11/113020}

\bibitem{Poyatos:1997}
Poyatos J, Cirac J and Zoller P 1997 {\em Physical Review Letters\/} {\bf 78}
  390--393 \urlprefix\url{http://link.aps.org/doi/10.1103/PhysRevLett.78.390}

\bibitem{Chuang:1997}
Chuang I~L and Nielsen M~A 1997 {\em Journal Of Modern Optics\/} {\bf 44}
  2455--2467
  \urlprefix\url{http://www.tandfonline.com/doi/abs/10.1080/09500349708231894}

\bibitem{Holtzafel2015}
Holz\"apfel M, Baumgratz T, Cramer M and Plenio M~B 2015 {\em Phys. Rev. A\/}
  {\bf 91}(4) 042129
  \urlprefix\url{https://link.aps.org/doi/10.1103/PhysRevA.91.042129}

\bibitem{Flammia2012}
Flammia S, Gross D, Liu Y~K and Eisert J 2012 {\em New Journal of Physics\/}
  {\bf 14} 095022
  \urlprefix\url{http://iopscience.iop.org/article/10.1088/1367-2630/14/9/095022/}

\bibitem{Flammia_adaptive}
Granade C, Ferrie C and Flammia S~T 2016 {\em arXiv:\/}  1605.05039

\bibitem{Merkel:2013}
Merkel S~T, Gambetta J~M, Smolin J~A, Poletto S, C{\'o}rcoles A~D, Johnson B~R,
  Ryan C~A and Steffen M 2013 {\em Physical Review A\/} {\bf 87} 062119
  \urlprefix\url{http://link.aps.org/doi/10.1103/PhysRevA.87.062119}

\bibitem{BlumeKohout:2017}
Blume-Kohout R, Gamble J~K, Nielsen E, Rudinger K, Mizrahi J, Fortier K and
  Maunz P 2017 {\em Nature Communications\/} {\bf 8} 14485
  \urlprefix\url{http://dx.doi.org/10.1038/ncomms14485}

\bibitem{Blatt_QPT}
Riebe M, Kim K, Schindler P, Monz T, Schmidt P~O, K\"orber T~K, H\"ansel W,
  H\"affner H, Roos C~F and Blatt R 2006 {\em Phys. Rev. Lett.\/} {\bf 97}(22)
  220407 \urlprefix\url{http://link.aps.org/doi/10.1103/PhysRevLett.97.220407}

\bibitem{Rutman}
Rutman J 1978 {\em Proc. IEEE\/} {\bf 66} 1048--1075

\bibitem{Hooge_1onf}
Hooge F~N, Kleinpenning T~G~M and Vandamme L~K~J 1981 {\em Reports on Progress
  in Physics\/} {\bf 44} 479
  \urlprefix\url{http://stacks.iop.org/0034-4885/44/i=5/a=001}

\bibitem{Clarke2004}
Harlingen D~J~V, Plourde B~L~T, Robertson T~L, Reichardt P~A and Clarke J 2004
  Decoherence in flux qubits due to $1/f$ noise in josephson junctions {\em
  Quantum Computing and Quantum Bits in Mesoscopic Systems\/} ed Leggett A,
  Ruggiero B and Silvestini P (New York: Kluwer Academic Press) pp 171--184

\bibitem{Dzurak_RB}
Fogarty M~A, Veldhorst M, Harper R, Yang C~H, Bartlett S~D, Flammia S~T and
  Dzurak A~S 2015 {\em Phys. Rev. A\/} {\bf 92}(2) 022326
  \urlprefix\url{http://link.aps.org/doi/10.1103/PhysRevA.92.022326}

\bibitem{Flammia_RBConfidence}
Wallman J~J and Flammia S~T 2014 {\em New Journal of Physics\/} {\bf 16} 103032
  \urlprefix\url{http://stacks.iop.org/1367-2630/16/i=10/a=103032}

\bibitem{Flammia_unitarity}
Wallman J, Granade C, Harper R and Flammia S~T 2015 {\em New Journal of
  Physics\/} {\bf 17} 113020
  \urlprefix\url{http://stacks.iop.org/1367-2630/17/i=11/a=113020}

\bibitem{Ball:2016}
Ball H, Stace T~M, Flammia S~T and Biercuk M~J 2016 {\em Physical Review A\/}
  {\bf 93} 022303
  \urlprefix\url{http://link.aps.org/doi/10.1103/PhysRevA.93.022303}

\bibitem{Flammia_FT}
Kueng R, Long D~M, Doherty A~C and Flammia S~T 2016 {\em Phys. Rev. Lett.\/}
  {\bf 117}(17) 170502
  \urlprefix\url{http://link.aps.org/doi/10.1103/PhysRevLett.117.170502}

\bibitem{Fong2017}
Fong B~H and Merkel S~T 2017 {\em arXiv:\/}  1703.09747

\bibitem{Wallman2017}
Wallman J~J 2017 {\em arXiv:\/}  1703.09835

\bibitem{Kueng:2016}
Kueng R, Long D~M, Doherty A~C and Flammia S~T 2016 {\em Physical Review
  Letters\/} {\bf 117} 170502
  \urlprefix\url{http://link.aps.org/doi/10.1103/PhysRevLett.117.170502}

\bibitem{Blume-Kohout2017}
Blume-Kahout R, Gamble J, Nielsen E, Mizrahi J, Fortier K and Maunz P 2017 {\em
  Nature Communications\/} {\bf 8} 14485

\bibitem{Ramsey1950}
Ramsey N~F 1950 {\em Phys. Rev.\/} {\bf 78}(6) 695--699
  \urlprefix\url{https://link.aps.org/doi/10.1103/PhysRev.78.695}

\bibitem{pyGSTi}
Nielsen E, Rudinger K, Gamble J~K and Blume-Kohout R 2016 pygsti: A python
  implementation of gate set tomography. available at http://github.com/pygstio

\bibitem{GST_Gauge}
Proctor T, Rudinger K, Young K, Sarovar M and Blume-Kohout R 2017 {\em
  arXiv:\/}  1702.01853

\bibitem{Rudnicki:2017}
Rudnicki {\L}, Pucha{\l}a Z and {\.{Z}}yczkowski K 2017 {\em arXiv.org\/}
  (\textit{Preprint} \eprint{1707.06926v1})
  \urlprefix\url{http://arxiv.org/abs/1707.06926v1}

\bibitem{Dehollain:2016}
Dehollain J~P, Muhonen J~T, Blume-Kohout R, Rudinger K~M, Gamble J~K, Nielsen
  E, Laucht A, Simmons S, Kalra R, Dzurak A~S and Morello A 2016 {\em New
  Journal of Physics\/} {\bf 18} 1--9
  \urlprefix\url{http://dx.doi.org/10.1088/1367-2630/18/10/103018}

\bibitem{Rudinger:2017}
Rudinger K, Kimmel S, Lobser D and Maunz P 2017 {\em Phys. Rev. Lett.\/} {\bf
  118}(19) 190502
  \urlprefix\url{https://link.aps.org/doi/10.1103/PhysRevLett.118.190502}

\bibitem{ENielsen:2017}
 private communications: {Erik Nielsen}

\end{thebibliography}

\end{document}